\documentclass[12pt]{article}
\usepackage{bbm,latexsym,amsmath}
\newcommand{\be}{\begin{equation}}
\newcommand{\ee}{\end{equation}}
\newcommand{\ba}{\begin{eqnarray}}
\newcommand{\ea}{\end{eqnarray}}

\renewcommand{\thefootnote}{\fnsymbol{footnote}}
\newcommand{\zz}{\mathbbm{Z}}
\newcommand{\mnu}{\mathcal{M}_\nu}

\newcommand{\diag}{\mbox{diag}\,}
\newcommand{\bone}{\mathbbm{1}}
\newtheorem{proposition}{Proposition}
\newtheorem{theorem}{Theorem}
\allowdisplaybreaks
\begin{document}
\renewcommand{\thefootnote}{\fnsymbol{footnote}}

\title{
\normalsize \hfill UWThPh-2012-36 \\[10mm]
\LARGE Discrete symmetries, roots of unity,\\ 
and lepton mixing}

\author{
W.~Grimus\thanks{E-mail: walter.grimus@univie.ac.at} 
\\[5mm]
\small University of Vienna, Faculty of Physics \\
\small Boltzmanngasse 5, A--1090 Vienna, Austria
\\[2mm]
}

\date{August 13, 2014}

\maketitle

\begin{abstract}
We investigate the possibility that
the first column of the lepton mixing matrix $U$ is given by 
$u_1 = (2,-1,-1)^T/\sqrt{6}$. In a purely group-theoretical approach, based
on residual symmetries in the charged-lepton and neutrino sectors and
on a theorem on vanishing sums of roots of unity, we discuss the
\emph{finite} groups which can enforce this. Assuming that there is
only one residual symmetry in the Majorana neutrino mass matrix, 
we find the  almost \emph{unique} solution 
$\zz_q \times S_4$ where the cyclic factor $\zz_q$ with $q = 1,2,3,\ldots$ 
is irrelevant for obtaining $u_1$ in $U$. Our discussion also provides
a natural mechanism for achieving this goal. Finally, 
barring vacuum alignment, we realize this
mechanism in a class of renormalizable models.
\end{abstract}

\newpage

\renewcommand{\thefootnote}{\arabic{footnote}}

\section{Introduction}
\label{intro}
The recent measurements of a rather large reactor mixing angle
$\theta_{13}$~\cite{daya-reno,schwetz} disfavour 
tri-bimaximal mixing~\cite{HPS} and, therefore, also such
models---see~\cite{altarelli,merlo} for reviews on models 
for neutrino masses and mixing. While the third column of the
tri-bimaximal mixing matrix $U_\mathrm{TBM}$ 
is now definitively in disagreement with
the data, the first or the second column of $U_\mathrm{TBM}$ 
could still occur in the mixing matrix $U$. These cases are denoted by TM$_1$
and TM$_2$, respectively, in~\cite{albright}.
However, in the case of TM$_2$ the solar mixing angle $\theta_{12}$ 
is related to the reactor mixing angle via 
$\sin^2 \theta_{12} (1 - \sin^2 \theta_{13}) = 1/3$, which creates a tension
with the data but is still compatible at the $3\sigma$ level. 
Therefore, it is more interesting to consider
TM$_1$~\cite{king,rodejohann,ma1,lavoura} where the first column in 
$U = \left( u_1, u_2, u_3 \right)$ is given by
\begin{equation}\label{TM1}
u_1 = \frac{1}{\sqrt{6}} \left( \begin{array}{c} 2 \\ -1 \\ -1 
\end{array} \right).
\end{equation}

Recently, a purely group-theoretical approach has been developed for the
investigation of the effect of \emph{finite} family symmetry groups on the
mixing matrix~\cite{lam,ge,yin,hagedorn,smirnov1,smirnov2,hu}.
Apart from assuming that the left-handed neutrino fields and the left-handed
charged lepton fields are in the same gauge doublet of the Standard Model
gauge group, no other assumption concerning the interactions in the lepton
sector is made. On the one hand, taking into account extant data on lepton
mixing, such a general approach allows a systematic
investigation of the possible symmetry groups, see for instance the scan of
groups performed in~\cite{lam1,lindner}. On the other hand, this 
approach has its limitations since it is not entirely clear how its results
relate to concrete models~\cite{grimus}; we will address this point
later in this paper. 

The goal of the present investigation is to find all possible finite family
symmetry groups underlying TM$_1$, without restricting the other two columns
$u_2$ and $u_3$ beyond orthonormality. 
In order to accomplish this task, we will
use a theorem on vanishing sums of roots of unity.
In the course of this investigation we will also come across a mechanism for
the implementation of TM$_1$. In the following we will use this mechanism to
write down a class of models based on $S_4$ and the type~II seesaw mechanism
where TM$_1$ is realized.

The paper is organized as follows. In section~\ref{residual} 
the group-theoretical method of~\cite{smirnov1} is reviewed using our
notation. In order to be as clear as possible, some arguments are emphasized
by formulating them as propositions. This also applies to
section~\ref{symmetry for TM1} where the symmetry group for 
TM$_1$ is determined; some technical points needed in the course of our
argumentation are deferred to appendix~\ref{roots of unity}. 
In the same section we also find a mechanism for the implementation of TM$_1$,
which is then used in section~\ref{concrete S_4 scheme} for the construction 
of a class of models. 
The summary of our findings is presented in section~\ref{summary}.
As a supporting material, we provide a set of generators of
$S_4$ and the three-dimensional irreducible representations of this group 
in appendix~\ref{generators}.

\section{Residual symmetries in the mass matrices}
\label{residual}
The class of models we have in mind as an application of the following
discussion are extensions of the Standard Model 
in the scalar and fermion sectors. Typical examples
would be several Higgs doublets and right-handed neutrino singlets
which facilitate the seesaw mechanism~\cite{seesaw}, or Higgs triplet
extensions with the type~II seesaw mechanism~\cite{II}. 
We further assume that before
spontaneous symmetry breaking (SSB) the theory is invariant under a 
\emph{finite} family symmetry group $G$ and that there are three 
lepton families. We want to investigate the
case that there are residual symmetries in the charged-lepton and
Majorana neutrino mass matrices, left over from SSB of $G$, and study
their effect on the lepton mixing matrix $U$~\cite{lam,smirnov1}. 

Such residual symmetries will occur whenever the vacuum expectation
values (VEVs) of the neutral components of scalar gauge multiplets are
invariant under some transformations of $G$. However, we are not
interested in the full symmetry of the vacuum, which might very well
be trivial, but in the symmetries of the vacua in the respective
sectors whose VEVs lead to charged-lepton and neutrino masses. It is well known
that the mismatch of these symmetries is responsible for predictions
in the mixing matrix---see for instance~\cite{altarelli,mismatch}. 
Therefore, we must distinguish between 
the symmetries in $M_\ell M_\ell^\dagger$ where $M_\ell$  
the charged-lepton mass matrix 
and those in the Majorana neutrino mass matrix
$\mnu$. These symmetries are supposed to 
generate the subgroups $G_\ell$ and $G_\nu$ of
$G$, pertaining to $M_\ell M_\ell^\dagger$ and $\mnu$, respectively. 
Obviously, we have the relation 
\begin{equation}\label{times}
G_\ell \subseteq U(1) \times U(1) \times U(1), \quad
G_\nu  \subseteq \zz_2 \times \zz_2 \times \zz_2
\end{equation}
due to the Dirac and Majorana natures of charged and neutral leptons,
respectively. The groups $G_\ell$ and $G_\nu$ will be quite small and
very often be generated by just one symmetry. 
In the following we will assume that this is the case for $G_\nu$, but
$G_\ell$ will in principle be allowed to contain several non-trivial
elements though the analysis will be phrased in terms of a 
single matrix $T \in G_\ell$.

The mass Lagrangian---obtained through SSB---has the form
\begin{equation}\label{mm}
\mathcal{L}_\mathrm{mass} = 
-\bar \ell_L M_\ell \ell_R + 
\frac{1}{2} \nu_L^T C^{-1} \mnu \nu_L + \mbox{H.c.} 
\end{equation}
with the residual symmetries 
\begin{equation}
T^\dagger M_\ell M_\ell^\dagger T = M_\ell M_\ell^\dagger, \quad
S^T \mnu S = \mnu,
\end{equation}
where $T$ and $S$ are unitary matrices and $\mnu$ is symmetric but
complex in general. Furthermore, the diagonalizing matrices $U_\ell$ and
$U_\nu$, and the lepton mixing matrix are given
\begin{equation}\label{UUU}
U_\ell^\dagger M_\ell M_\ell^\dagger U_\ell = 
\diag \left( m_e^2, m_\mu^2, m_\tau^2 \right), \quad
U_\nu^T \mnu U_\nu = \diag \left( m_1, m_2, m_3 \right), \quad 
U = U_\ell^\dagger U_\nu,
\end{equation}
respectively.
We denote the weak basis of $M_\ell$ and $\mnu$ by basis~1.
Since we are interested in the effect of $T$ and $S$ on the lepton
mixing matrix, the action of the residual symmetry on right-handed lepton
fields is irrelevant for us, which is the reason to consider 
$M_\ell M_\ell^\dagger$ instead of $M_\ell$. 

It is useful to distinguish basis~1 from the weak basis,
where the charged-lepton mass matrix is diagonal, in which case we use the
phrase basis~2. All matrices in this basis are indicated by a tilde. Usually, 
basis~1 is the weak basis where the matrices of the representation $D(G)$ on
the left-handed lepton gauge doublets have a ``nice'' form.
Depending on the representations of $G$ used in the Lagrangian it can
happen that basis~1 coincides with basis~2, but in general the two
bases will be different. Because all charged-lepton masses are
different, we conclude that $\tilde T$ is a diagonal matrix.

In the following analysis 
we will assume that there are no accidental symmetries
in $M_\ell M_\ell^\dagger$ and $\mnu$. 
Then, since $S$ and $T$ are given in a weak basis, they must both belong to
$D(G)$, the group of representation matrices of $G$ acting on the left-handed
lepton doublets. 
The matrices $S$ and $T$ generate a group denoted by 
$\bar G$. In this case it is clear that $\bar G$
is a subgroup of the group $D(G)$ and both are subgroups of $U(3)$, i.e.\  
$\bar G \subseteq D(G) \subset U(3)$.
In the simplest case $\bar G$ is identical with $D(G)$.

Due to equation~(\ref{times}) and assuming $\det S = 1$ without loss
of generality, the form of $S$ is given by
\begin{equation}\label{inv}
S = 2uu^\dagger - \bone
\end{equation}
with a unit vector $u$ and $S^2 = \bone$.
\begin{proposition}\label{1}
If $S^T \mnu S = \mnu$ with $S = 2uu^\dagger - \bone$, 
then $\mnu u \propto u^*$.
\end{proposition}
Proof: By construction, the matrix $S$ fulfills $Su = u$, with a
unique eigenvalue $1$. Therefore, $S^T \mnu u = \mnu u$, and due to the
hermiticity of $S$ and $S^2 = \bone$ we find $S^* (\mnu u) = \mnu u$. 
Since $S^*$ has a unique eigenvalue $1$ with corresponding eigenvector
$u^*$, we conclude that $\mnu u$ is proportional to $u^*$. $\Box$

In basis~2 the mixing matrix $U$ diagonalizes the neutrino mass matrix:
\begin{equation}\label{U}
U^T \widetilde \mnu U = \diag (m_1, m_2, m_3).
\end{equation}
With column vectors $u_j$ and $U = (u_1,u_2,u_3)$, equation~(\ref{U})
is reformulated as 
\begin{equation}
\widetilde \mnu u_j = m_j u_j^*.
\end{equation}
Comparing with proposition~\ref{1} and denoting the unit vector associated
with $\tilde S$ by $\tilde u$, 
we immediately come to the following conclusion.
\begin{proposition}\label{2}
If ${\tilde S}^T \widetilde \mnu \tilde S = \widetilde \mnu$, then,
apart from irrelevant phases,
$\tilde u$ is one of column vectors of the lepton mixing matrix $U$.
\end{proposition}
This proposition is the basis of all discussions concerning residual
symmetries in the mass matrices.

Since we assume that $G$ is a finite group, it is a necessary condition that 
the matrices $T$ and $ST$ have finite orders:
\begin{equation}\label{mn}
T^m = (ST)^n = \bone
\end{equation}
for some natural numbers $m$ and $n$. Note that $S$ has
order two by construction.\footnote{Groups generated by $S$ and $T$ such that
  the orders of $S$, $T$ and $ST$ are finite are called von Dyck
  groups~\cite{smirnov1}; such groups are not necessarily finite because
  because finite orders of group generators does in general not imply that the
  group is finite.}

The interesting observation is that  
equation~(\ref{mn}) connects group properties with information on the
mixing matrix~\cite{lam,smirnov1}.
In effect, one uses the following proposition.
\begin{proposition}
Suppose we have a $3 \times 3$ matrix $M$ which is a function of a set
of parameters $x = (x_1, \ldots, x_r)$ such that for all values of $x$
$M$ is unitary. Then the values of $x$ where $M$ has the
eigenvalues $\lambda_k$ ($k=1,2,3$ and $|\lambda_k| = 1$) are
determined by the two equations 
\begin{equation}\label{tr-det}
\mathrm{Tr}\, M(x) = \lambda_1 + \lambda_2 + \lambda_3 
\quad \mbox{and} \quad 
\det M(x) = \lambda_1 \lambda_2 \lambda_3.
\end{equation}
\end{proposition}
Proof: The characteristic polynomial of $M$ is 
\[
P_M(\lambda) = \lambda^3 - M_2 \lambda^2 + M_1 \lambda - M_0.
\]
In terms of the eigenvalues of $M$ the coefficients $M_i$ 
are given by 
\[
M_1 = \lambda_1 \lambda_2 + \lambda_2 \lambda_3 + \lambda_3 \lambda_1,
\quad
M_2 = \lambda_1 + \lambda_2 + \lambda_3, 
\quad
M_0 = \lambda_1 \lambda_2 \lambda_3.
\]
Unitarity of $M$ means that the eigenvalues are located on the unit
circle. Therefore, we find $M_1 = M_2^* M_0$. This means that, as long
as equation~(\ref{tr-det}) is satisfied, then automatically $M$ gives
the correct coefficient $M_2$ in $P_M(\lambda)$ leading to the
eigenvalues $\lambda_k$. $\Box$ 
\\[1mm] \noindent
We make the identification $M = \tilde S \tilde T$. With  
\begin{equation}
\tilde T = \diag \left( e^{i\phi_e}, e^{i\phi_\mu}, e^{i\phi_\tau} \right)
\quad \mbox{and} \quad
\tilde u = \left( \begin{array}{c} 
U_{ei} \\U_{\mu i} \\U_{\tau i} 
\end{array} \right)
\end{equation}
equation~(\ref{tr-det}) reads~\cite{smirnov1}
\begin{equation}\label{master}
\sum_{\alpha = e,\mu,\tau} \left( 2\left| U_{\alpha i} \right|^2 - 1
\right) e^{i\phi_\alpha} = \lambda_1 + \lambda_2 + \lambda_3
\quad \mbox{and} \quad
\prod_\alpha e^{i\phi_\alpha} = \lambda_1 \lambda_2 \lambda_3,
\end{equation}
where the $e^{i\phi_\alpha}$ are $m$-th roots and the $\lambda_k$ are
$n$-th roots of unity. The parameters in $M$ are the mixing angles, 
the CP phase and the $\phi_\alpha$. 
Equation~(\ref{master}) provides the necessary and sufficient
conditions that $ST$ is of order~$n$.

Equation~(\ref{master}) can be used both ways: Given the group $\bar G$
generated by $S$ and $T$, relations among the mixing parameters are
obtained~\cite{smirnov1}; vice versa, assuming a specific column vector
in $U$, we can infer the group $\bar G$.

\section{A symmetry for TM$_1$}
\label{symmetry for TM1}
Now we discuss the possible symmetry groups $\bar G$ leading to TM$_1$.
From the discussion in the previous section we have learned 
that in this case we have to take $\tilde u \equiv u_1$ with $u_1$ given by
equation~(\ref{TM1}), which leads to 
\begin{equation}\label{Stilde}
\tilde S = \frac{1}{3} \left( \begin{array}{rrr}
1 & -2 & -2 \\ -2 & -2 & 1 \\ -2 & 1 & -2
\end{array} \right).
\end{equation}
Specifying equation~(\ref{master}) to this case, we obtain
\begin{equation}\label{master1}
-e^{i \phi_e} + 2 e^{i\phi_\mu} + 2 e^{i\phi_\tau} + 3 \lambda_1 +  
3 \lambda_2 +  3 \lambda_3 = 0
\quad \mbox{and} \quad
e^{i\phi_e} e^{i\phi_\mu} e^{i\phi_\tau} = \lambda_1 \lambda_2 \lambda_3.
\end{equation}
Our aim is to find all cases of roots of unity 
$e^{i\phi_\alpha}$ ($\alpha = e,\mu,\tau$) and $\lambda_k$ ($k=1,2,3$)
which satisfy these two relations. For this purpose we use a
theorem by Conway and Jones, theorem~6 in~\cite{conway}.

Some remarks are in order before we reproduce this theorem.
Formal sums of roots of unity with rational coefficients form
a ring~\cite{conway}. 
A sum of roots of unity $\mathcal{S}'$ is \emph{similar} to 
a sum of roots of unity $\mathcal{S}$
if there is a rational number $q$ and a root of unity
$\delta$ such that $\mathcal{S}' = q \delta \mathcal{S}$.
The length of a sum of roots of unity $\mathcal{S}$ is the number of
roots involved. Note, however, that, for the definition of this
length, roots of unity which differ only by a sign do not count separately.
The roots occurring in the following theorem are
\begin{equation}\label{obg}
\omega = e^{2\pi i/3}, \quad 
\beta = e^{2\pi i/5}, \quad
\gamma = e^{2\pi i/7}. 
\end{equation}
\begin{theorem}[Conway and Jones]\label{theorem}
Let $\mathcal{S}$ be a non-empty vanishing sum of length at most~9. 
Then either $\mathcal{S}$ involves $\theta$, $\theta\omega$, 
$\theta\omega^2$ for some root $\theta$, or $\mathcal{S}$ is similar
to one of 
\begin{enumerate}
\renewcommand{\labelenumi}{\alph{enumi})}
\item
$1 + \beta + \beta^2 +\beta^3 +\beta^4$,
\item 
$- \omega - \omega^2 + \beta + \beta^2 +\beta^3 +\beta^4$, 
\item
$1 + \gamma + \gamma^2 +\gamma^3 +\gamma^4 +\gamma^5 +\gamma^6$, 
\item
$1 + \beta + \beta^4 - (\omega + \omega^2)(\beta^2 + \beta^3)$,
\item
$- \omega - \omega^2 + \gamma + \gamma^2 +\gamma^3 +\gamma^4
  + \gamma^5 + \gamma^6$,
\item 
$\beta + \beta^4 - (\omega + \omega^2)(1 + \beta^2 + \beta^3)$,
\item
$1 + \gamma^2 +\gamma^3 +\gamma^4 +\gamma^5 -( \omega +
  \omega^2)(\gamma + \gamma^6)$,
\item 
$1 - (\omega + \omega^2)(\beta + \beta^2 +\beta^3 +\beta^4)$.
\end{enumerate}
\end{theorem}

The first relation of equation~(\ref{master1}) constitutes a vanishing
sum $\mathcal{S}$ whose length is at most~6. It is immediately obvious that the
vanishing sums in the theorem cannot be part of $\mathcal{S}$, because
c)--h) are too long, sum~b) is not similar to $\mathcal{S}$ and sum~a)
is a vanishing sum of length~5 and adding one root would
make a non-vanishing sum. Therefore, in the case
of~equation~(\ref{master1}), the theorem says that $\mathcal{S}$ must
involve $\omega$. Exploiting this fact in 
appendix~\ref{roots of unity}, we find that equation~(\ref{master1})
has the solution 
\begin{equation}\label{solution}
e^{i \phi_e} = \eta, \;\;
e^{i\phi_\mu} = \eta \omega, \;\; 
e^{i\phi_\tau} =\eta \omega^2, \;\; 
\lambda_1 = \epsilon,\;\;
\lambda_2 = -\epsilon,\;\;
\lambda_3 = \eta,
\end{equation}
with arbitrary roots of unity $\epsilon$ and $\eta$.
The second condition of equation~(\ref{master1}) yields
\begin{equation}
\eta^3 = -\epsilon^2 \eta \quad \mbox{or} \quad
\epsilon = \pm i \eta.
\end{equation}
However, this equation will be of no use in the following. 

In summary, our considerations has lead us to
\begin{equation} \label{Ttilde}
\tilde T = \eta\, \diag \left( 1, \omega, \omega^2 \right).
\end{equation}
Thus $\tilde T$ is completely determined up to an unknown root of unity
$\eta$. 
We remind the reader that 
the matrices $\tilde S$ of equation~(\ref{Stilde}) and $\tilde T$ of
equation~(\ref{Ttilde}) are given in basis~2. 

A suitable basis~1---see appendix~\ref{generators}---is given by a basis
transformation with
\begin{equation}\label{Uomega}
U_\omega = \frac{1}{\sqrt{3}} \left(
\begin{array}{ccc}
1 & 1 & 1 \\ 1 & \omega & \omega^2 \\ 1 & \omega^2 & \omega
\end{array} \right).
\end{equation} 
The resulting matrices are 
\begin{equation}\label{ST}
S = U_\omega \tilde S U_\omega^\dagger = S_1 B =
\left(\!\! \begin{array}{rcc}
-1 & 0 & 0 \\ 0 & 0 & 1 \\ 0 & 1 & 0 
\end{array} \right)
\quad \mbox{and} \quad 
T = U_\omega \tilde T U_\omega^\dagger = \eta E =
\eta \left( \begin{array}{rrr}
0 & 1 & 0 \\ 0 & 0 & 1 \\ 1 & 0 & 0 
\end{array} \right),
\end{equation}
where $S_1$ is defined in equation~(\ref{Si}) and $E$ and $B$ in
equation~(\ref{EB}). 

Let us assume that $\eta$ is a primitive root
of order $q$. The following proposition determines the group $\bar G$.
\begin{proposition}\label{s4}
The matrix $\tilde S$ of equation~(\ref{Stilde}) and $\tilde T$ of
equation~(\ref{Ttilde}) generate the group $\bar G = \zz_q \times S_4$.
\end{proposition}
Proof: For simplicity of notation we use basis~1. From 
$(ST)^2 = \eta^2 S_3$ and $T^\dagger \eta^2 S_3 T = \eta^2 S_1$ 
we find that $\eta^2 S_1 S = \eta^2 B \in \bar G$ and, therefore,  
\[
(\eta^2 B)^\dagger T (\eta^2 B) = 
\eta\,\diag \left( 1,\omega^2,\omega \right) \in \bar G.
\]
Eventually, with 
\[
T (\eta^2 B)^\dagger T (\eta^2 B) = \eta^2 \bone \in \bar G
\quad \mbox{and} \quad
T^3 = \eta^3 \bone \in \bar G
\]
we conclude $\eta \bone \in \bar G$. Therefore, $\bar G$ contains $E$, $S_1$
and $B$, which is a set of generators of $S_4$---see appendix~\ref{generators}. 
It is then almost trivial to show that every element of $g \in \bar G$ 
can uniquely be decomposed into $g = \eta^k h$ with 
$k \in \{ 0, 1, \ldots, q-1 \}$ and $h \in S_4$. $\Box$

In summary, we have found \emph{all} finite groups which enforce TM$_1$ in the
mixing matrix because we have seen that any $\tilde T$ of finite order
satisfying equation~(\ref{master1}) leads to $S_4$ times a cyclic factor. 
In~\cite{lam1} it was stated that $S_4$ is the smallest
such group. Here we have shown that actually $S_4$ is unique up to a
trivial factor with a cyclic group. We formulate our result as a theorem.
\begin{theorem}\label{main}
Under the premises that $G_\nu$ is a $\zz_2$ generated by $\tilde S$ of
equation~(\ref{Stilde}) and that $G_\ell$ contains at least one
matrix $\tilde T$ which is not proportional to the unit
matrix, the only symmetry groups $\bar G$ 
generated by the residual symmetries of the mass
matrices which are able to enforce TM$_1$ in the lepton mixing matrix $U$ 
are $\zz_q \times S_4$.
\end{theorem}

With our derivation of theorem~\ref{main} we have also demonstrated that
\begin{equation}\label{ME}
E^\dagger \left( M_\ell M_\ell^\dagger \right) E = 
M_\ell M_\ell^\dagger.
\end{equation}
This has the following consequence.
\begin{proposition}
For every charged-lepton mass matrix which fulfills
equation~(\ref{ME}) it follows that
$U_\omega^\dagger \left( M_\ell M_\ell^\dagger \right) U_\omega$ is diagonal.
\end{proposition}
Proof: The eigenvectors of $E$ are identical with the column vectors
of $U_\omega$, i.e.\ $U_\omega = \left( x_1,x_2,x_3 \right)$ and 
$E x_k = \omega^{k-1} x_k$. Therefore, we have
\[
\left( M_\ell M_\ell^\dagger \right) E x_k = 
\omega^{k-1} \left( M_\ell M_\ell^\dagger \right) x_k =
E \left( M_\ell M_\ell^\dagger \right) x_k,
\]
whence we conclude that $\left( M_\ell M_\ell^\dagger \right) x_k$ is
an eigenvector of $E$ to the eigenvalue $\omega^{k-1}$. Since
eigenvectors to non-degenerate eigenvalues are unique up to
a multiplicative factor, we arrive at 
$\left( M_\ell M_\ell^\dagger \right) x_k = \nu_k x_k$ where the
quantities $|\nu_k|$ are identical with the charged-lepton masses. $\Box$

Turning to the neutrino mass matrix $\mnu$, we remember that we have started
with the requirement that it is invariant under $S$, i.e.\ 
$S^T \mnu S = \mnu$. In the present discussion $S$
is given by equation~(\ref{ST}). 
By construction, this matrix is an involution---see
also equation~(\ref{inv})---with a unique eigenvalue~$1$:
\begin{equation}\label{u}
Su = u = U_\omega u_1 \quad \mbox{with} \quad 
u = \frac{1}{\sqrt{2}} \left( \begin{array}{c} 
0 \\ 1 \\ 1 
\end{array} \right).
\end{equation}
Then we know from proposition~\ref{1} that $u$ is also an
eigenvector of $\mnu$. Therefore, in our basis~1 the mechanism for
achieving TM$_1$ boils down to
\begin{equation}\label{uu}
U_\omega^\dagger u = 
\frac{1}{\sqrt{6}} \left( \begin{array}{c} 2 \\ -1 \\ -1 
\end{array} \right).
\end{equation}
This mechanism has recently been used in~\cite{lavoura} for the construction
of a model which exhibits TM$_1$.

\section{Realizing TM$_1$ in a concrete $S_4$ scheme}
\label{concrete S_4 scheme}
Though the mechanism for obtaining TM$_1$ from the mass matrices is unique,
it is not unique how to embed it into an $S_4$ model. 
Below we introduce a scheme with the type~II seesaw mechanism~\cite{II}.

Our starting point is the tensor product (see for instance~\cite{GL-review})
\begin{equation}\label{33}
{\bf 3} \otimes {\bf 3} = 
{\bf 1} \oplus {\bf 2} \oplus {\bf 3} \oplus {\bf 3'},
\end{equation}
where the ${\bf 1}$ is the trivial one-dimensional representation and the 
{\bf 3} and {\bf 3'} are the two inequivalent irreducible three-dimensional
representations of $S_4$---see  appendix~\ref{generators} for the generators
and the three-dimensional representations.
The ${\bf 3'}$ and ${\bf 3}$ correspond to the off-diagonal symmetric and
antisymmetric parts, respectively, in the tensor product.
The ${\bf 1}$ and the ${\bf 2}$ comprise the diagonal part.
If we assign to both the left-handed lepton gauge doublets and the
right-handed lepton gauge singlets a ${\bf 3}$ of $S_4$, then the
right-hand side of equation~(\ref{33}) shows the possible 
irreducible $S_4$ representations of Higgs doublets. 

Since we require the validity of equation~(\ref{ME}), we need the
vacuum to be invariant under $s = (123)$, which is mapped in the 
${\bf 3}$ and ${\bf 3'}$ into $E$---see appendix~\ref{generators}.  
However, for the two-dimensional irreducible representation we have 
\begin{equation}
s = (123) \mapsto 
\left( \begin{array}{cc} \omega & 0 \\ 0 & \omega^2
\end{array} \right),
\end{equation}
which means that no non-trivial VEV is invariant under this
representation of $s$ and the ${\bf 2}$ cannot contribute to $M_\ell$.
For ${\bf 3}$ and ${\bf 3'}$, invariance of the VEVs under $s$ means
that the VEVs of the three Higgs doublets have to be equal.
Eventually, we arrive at the most general $M_\ell$, invariant under
$E$ and compatible with equation~(\ref{33}):
\begin{equation}\label{ml}
M_\ell = \left( \begin{array}{ccc}
a & b+c & b-c \\ b-c & a & b+c \\ b+c & b-c & a
\end{array} \right).
\end{equation}
Indeed, here $U_\omega$ diagonalizes not only
$M_\ell M_\ell^\dagger$ but also $M_\ell$:
\begin{equation}
U_\omega^\dagger M_\ell U_\omega = \diag \left(
a + 2b,\, a - b + \sqrt{3} i c,\, a - b - \sqrt{3} i c
\right).
\end{equation}
This result shows that $M_\ell$ of equation~(\ref{ml}) is rich enough 
to accommodate three different charged-lepton masses, albeit with finetuning.

Using equation~(\ref{33}) in the neutrino sector, we note that 
the antisymmetric part on the right-hand side, the ${\bf 3}$, is not
allowed according to the assumed Majorana nature of the neutrinos.
Dropping also the ${\bf 2}$, 
we have four scalar gauge triplets $\Delta_k$ ($k=0,1,2,3$) in 
${\bf 1} \oplus {\bf 3'}$. 
From equation~(\ref{ST}) we know that $\mnu$ has to be invariant under
$S_1B$. Therefore, the triplet VEVs $w_k$ ($k=0,1,2,3$) 
have to be invariant under the $S_4$ transformation corresponding to
$S_1B$ in the representation of the scalar triplets. 
Leaving out the trivial case of $w_0$ and using that $S_1B$ acts as
$-S_1B$ on the ${\bf 3'}$, 
the VEVs $w_k $ ($k=1,2,3$) are determined by 
\begin{equation}
-S_1B \left( \begin{array}{c} w_1 \\ w_2 \\ w_3 
\end{array} \right) = 
\left( \begin{array}{c} w_1 \\ w_2 \\ w_3 
\end{array} \right) 
\quad \Rightarrow \quad w_3 = -w_2,
\end{equation}
and, therefore, the neutrino mass matrix has the form
\begin{equation}
\mnu = \left( \begin{array}{ccc}
A & B & -B \\ B & A & C \\ -B & C & A
\end{array} \right).
\end{equation}
The mass matrices $M_\ell$ and $\mnu$ of this section have
recently been obtained in~\cite{lavoura} in a different model.

\section{Summary}
\label{summary}

Before we summarize our findings, we want to point out the caveats and
limitations attached to the group-theoretical method reviewed in
section~\ref{residual}. It is useful to distinguish between three groups: $G$
is the family symmetry group of the Lagrangian, 
the group $D(G)$ is the $U(3)$ subgroup given by the representation
matrices of $G$ on the three left-handed leptonic gauge doublets,
and $\bar G$ is the  $U(3)$ subgroup generated by the residual symmetries of
the mass matrices $M_\ell M_\ell^\dagger$ and $\mnu$.
The method of section~\ref{residual} is a prescription for the 
determination of $\bar G$. 
How $\bar G$ is related to $G$ in a specific model and what $\bar G$ tells us
about model building, is another matter. This always has to be kept in mind
when assessing results obtained by the group-theoretical method of
section~\ref{residual}. Below, whenever we use
the phrase ``mass matrices,'' we mean $M_\ell M_\ell^\dagger$ and $\mnu$.
Our list of caveats is the following:
\begin{itemize}
\item
The method of section~\ref{residual} explicitly assumes that the family
symmetry group $G$ of the Lagrangian is finite and that neutrinos have
Majorana nature. 
\item
Since this method is purely group-theoretical and uses only information
contained in the mass matrices, it can yield at most $D(G)$.
\item
It is well known that accidental symmetries can occur in the mass matrices,
which contribute, therefore, to $\bar G$.
An accidental symmetry cannot be elevated to a
symmetry of the Lagrangian. 
In this case, $\bar G$ is not even a subgroup of $D(G)$---see for
instance~\cite{grimus}. 
\item
The method does not apply to models where VEVs break $G$ totally.
\end{itemize}
Note that it is possible that a model is predictive because of an
accidental symmetry.\footnote{The typical $A_4$ models~\cite{altarelli} have a
  $\zz_2 \times \zz_2$ symmetry in $\mnu$, but only one $\zz_2$ is the residual
  symmetry of $A_4$.} 
Even if $G$ is totally broken and there are no accidental
symmetries, the model can be predictive because 
of the restrictions imposed by $G$ on the Yukawa couplings or because the
VEVs have an alignment but this does not correspond to a subgroup of $G$.
For instance, the typical neutrino mass matrix resulting from
$\Delta(27)$ is a case where the group is completely broken, but has
a predictive neutrino mass matrix in specific models~\cite{ma}.

The methods and results of the paper can be summarized as follows:
\begin{enumerate}
\renewcommand{\labelenumi}{\roman{enumi})}
\item
We have used the group-theoretical method of~\cite{smirnov1}, together with
theorem~\ref{theorem} on vanishing sums of roots of unity, to determine all
possible groups $\bar G$ which result from the requirement that in the lepton
mixing matrix $U$ the first column is given by equation~(\ref{TM1}),
i.e.\ identical with the first column of the tri-bimaximal mixing matrix.
This is called TM$_1$ in~\cite{albright}.
\item
The result is amazingly simple. Only groups $\bar G$ of the form 
$\zz_q \times S_4$ with $q = 1,2,3, \ldots$ are capable to enforce TM$_1$
without fixing the columns $u_2$ and $u_3$ in $U$.
Note that we have not only shown that all such groups contain
$S_4$~\cite{lam1}, from our investigation it follows that any such group 
larger than $S_4$ is obtained from $S_4$ by multiplication with a cyclic
factor.\footnote{In~\cite{lam} $S_4$ was identified as the minimal
  group $\bar G$ for tri-bimaximal mixing, but it is also the group for TM$_1$
  alone~\cite{lam1,grimus}; this follows for instance from
  proposition~\ref{s4}. The reason is that $S_4$ contains not only 
  $\tilde S_1$ of equation~(\ref{Stilde}) but also 
  $\tilde S_i = 2 u_i u_i^\dagger - \bone$ ($i=2,3$) 
  where $u_2$ and $u_3$ are the second and third column of $U_\mathrm{TBM}$,
  respectively. When 
  $S_4$ is the group of TM$_1$, then $\tilde S_i$ with $i=2,3$ is broken and
  not part of $G_\nu$.}  
\item
Furthermore, in the basis where the Klein four-group, which is a subgroup of
$S_4$---see appendix~\ref{generators}, 
is represented by diagonal matrices, we have found a unique mechanism
for achieving TM$_1$: the first column in $U_\nu$ must be the vector $u$ of
equation~(\ref{u}), while $U_\ell = U_\omega$---see equation~(\ref{Uomega}). 
This mechanism was recently used in~\cite{lavoura}.
\item
Finally, we have pointed out how to straightforwardly 
implement the mechanism of the previous
item in a class of renormalizable $S_4$ models with type~II seesaw
mechanism. It is fair to mention that we have not solved the VEV alignment
problem in this context. 
\end{enumerate}

\vspace{5mm}\noindent
\textbf{Acknowledgments:} The author thanks P.O.~Ludl for many helpful
discussions and L.~Lavoura for comments on an early version of the manuscript.
Moreover, the author is very much indebted to R.M. Fonseca for
pointing out the erroneous theorem~1 in the previous version of the
paper, which has been replaced here by the theorem of Conway and Jones.

\appendix

\setcounter{equation}{0}
\renewcommand{\theequation}{A\arabic{equation}}

\section{Roots of unity and the eigenvalues of $T$ and $ST$}
\label{roots of unity}
Here we discuss the general solution of 
\begin{equation}\label{master2}
\mathcal{S} \equiv 
-e^{i \phi_e} + 2 e^{i\phi_\mu} + 2 e^{i\phi_\tau} + 3 \lambda_1 +  
3 \lambda_2 +  3 \lambda_3 = 0,
\end{equation}
where $e^{i \phi_\alpha}$ ($\alpha = e,\mu,\tau$) and 
$\lambda_k$ ($k=1,2,3$) are roots of unity, \textit{c.f.}\ 
equation~(\ref{master1}). 
As argued in section~\ref{symmetry for TM1}, 
according to theorem~\ref{theorem} (theorem~6 of~\cite{conway}), any
solution of this equation can only involve partial sums  
$\theta_j \left( 1 + \omega + \omega^2 \right)$ and ``empty sums''
$\epsilon_\kappa - \epsilon_\kappa$ 
for some roots of unity $\theta_j$ and $\epsilon_\kappa$.
$\mathcal{S}$ has 14 roots, though at most six of them
are different. Therefore, $\mathcal{S}$ must have one of the following 
three forms:
\begin{equation}\label{i}
\mathcal{S} =
\sum_{j=1}^4 \theta_j \left( 1 + \omega + \omega^2 \right) + 
\epsilon_1 - \epsilon_1 
\end{equation}
or
\begin{equation}\label{ii}
\mathcal{S} =
\sum_{j=1}^2 \theta_j \left( 1 + \omega + \omega^2 \right) + 
\sum_{\kappa=1}^4 \left( \epsilon_\kappa - \epsilon_\kappa \right) 
\end{equation}
or
\begin{equation}\label{iii}
\mathcal{S}  =
\sum_{\kappa=1}^7 \left( \epsilon_\kappa - \epsilon_\kappa \right).
\end{equation}
Clearly, in order to obtain solutions of equation~(\ref{master2}),
we have to reduce the number of different roots occurring in
equations~(\ref{i})--(\ref{iii}) to at most six. 
To facilitate this task, we define the sets
$M_j = \{ \theta_j, \theta_j \omega, \theta_j \omega^2 \}$ and observe
that for $j \neq j'$ it follows that either $M_j = M_{j'}$ or
$M_j \cap M_{j'} = \emptyset$. Furthermore, in a sum of roots of
unity, we call the
\emph{positive} coefficients in front of the roots weights. 
Thus in equation~(\ref{master2}) the weights 
are $3,\,3,\,3,\,2,\,2$ and~$1$, 
with the sum over the weights being 14.

First we discuss equation~(\ref{i}). In order to reduce the number of
different roots in $\mathcal{S}$ to at most six, without loss of generality we
have to make one of the following identifications:
\[
M_1 = M_2 = M_3 = M_4, \quad \mbox{or} \quad
M_1 = M_2 = M_3 \neq M_4, \quad \mbox{or} \quad
M_1 = M_2 \neq M_3 = M_4.
\]
In the first case we have 
$
\mathcal{S} = 4\theta_1 \left( 1 + \omega + \omega^2 \right) + 
\epsilon_1 - \epsilon_1.
$
In order to avoid two weights~1 in $\mathcal{S}$, we have to identify
either $\epsilon_1$ or $-\epsilon_1$ with a root in $M_1$. Without
loss of generality we put $\epsilon_1 = \theta_1$ and obtain 
$\mathcal{S} = 5\theta_1 + 4\theta_1 \omega + 4\theta_1 \omega^2 - \theta_1$.
It is then obvious that we cannot achieve three weights~3 and two weights~2. 
Regarding the second and third case, 
in order to avoid more than six different roots,
we must require $\epsilon_1 \in M_2$ and $-\epsilon_1 \in M_4$;
without loss of generality, this requirement is satisfied by
$\theta_1 = -\theta_4$. 
But then the second case leads to
$\mathcal{S} = 3\theta_1 \left( 1 + \omega + \omega^2 \right)
-\theta_1 \left( 1 + \omega + \omega^2 \right) + \theta_1 - \theta_1$.
Indeed there are three weights~3, but not two weights~2.
In the third case we have
$
\mathcal{S} = 2\theta_1 \left( 1 + \omega + \omega^2 \right)
-2\theta_1 \left( 1 + \omega + \omega^2 \right) + \theta_1 - \theta_1$.
Now there are two weights~3 instead of three.
In summary, we find that equation~(\ref{i}) cannot lead to a solution
of equation~(\ref{master2}). 

Next we consider equation~(\ref{iii}).
Without loss of generality, in order to produce three weights~3, 
we make the identifications 
$\epsilon_1 = \epsilon_2 = \epsilon_3$ and 
$\epsilon_4 = \epsilon_5 = \epsilon_6$. Thus we consider
$\mathcal{S} = 
3 \left( \epsilon_1 - \epsilon_1 +  \epsilon_4 - \epsilon_4 \right) +
\epsilon_7 - \epsilon_7.$
In order to avoid two weights~1, we furthermore equate, without loss
of generality, $\epsilon_7$ with $\epsilon_4$.
In this way, we arrive at
$\mathcal{S} = 
3 \left( \epsilon_1 - \epsilon_1 \right) +  
4 \left( \epsilon_4 - \epsilon_4 \right).
$
Now it suffices to consider only the non-trivial case 
$\epsilon_4 \neq \pm \epsilon_1$. It is then clear that the resulting 
$\mathcal{S}$ leads to a solution
of equation~(\ref{master2}) given by
$\lambda_1 = -\lambda_2 = \epsilon_1$, $\lambda_3 = \mp\epsilon_4$, 
$e^{i\phi_e} = e^{i\phi_\mu} = e^{i\phi_\tau} = \pm \epsilon_4$ and 
$T = \pm \epsilon_4 \bone$.
But this $T$ is trivial and we discard equation~(\ref{iii}) as well.

Considering equation~(\ref{ii}), we have either 
$M_1 = M_2$ or $M_1 \neq M_2$. We begin with the second case. 
In order to have at most six different roots, without loss of generality, 
we assume $\epsilon_\kappa \in M_1$ $\forall\,\kappa=1,\ldots,4$,
which necessitates $\theta_2 = -\theta_1$.
In this case, for every root in $M_1$
there is a corresponding root with the opposite sign in
$M_2$. Consequently, all weights occur in even numbers, which is a
contradiction to equation~(\ref{master2}). Thus we are left with 
$M_1 = M_2$.
We first envisage the possibility that the $\epsilon_\kappa$ 
do \emph{not} produce weight~3. 
Then, in order to obtain three weights~3, we identify
$\epsilon_\kappa$ with $\theta_1 \omega^{\kappa-1}$ for 
$\kappa=1,2,3$, which yields the sum  
$\mathcal{S} = 3 \theta_1 \left( 1 + \omega + \omega^2 \right) - %
\theta_1 \left( 1 + \omega + \omega^2 \right) + %
\epsilon_4 - \epsilon_4$.
Now there is no choice of $\epsilon_4$ such that the weights of
equation~(\ref{master2}) are reproduced. 
It remains to consider the case that the $\epsilon_\kappa$ do produce 
weight~3, \textit{i.e.}\ without loss of generality we assume
$\epsilon_1 = \epsilon_2 = \epsilon_3$. With this we obtain the sum 
$\mathcal{S} = 2\theta_1 \left( 1 + \omega + \omega^2 \right) + %
3 \left( \epsilon_1 - \epsilon_1 \right) + \epsilon_4 - \epsilon_4$.
Since we need a third weight~3, without loss of generality we make the
identification $\epsilon_4 = \theta_1$. Now we have indeed a viable
solution of equation~(\ref{master2}), given by
$\lambda_1 = -\lambda_2 = \epsilon_1$, $\lambda_3 = \theta_1$, 
$e^{i\phi_e} = \theta_1$, 
$e^{i\phi_\mu} = \theta_1 \omega$,
$e^{i\phi_\tau} = \theta_1 \omega^2$. 
With $\theta_1 \equiv \eta$, $\epsilon_1 \equiv \epsilon$ this is
precisely the solution presented in equation~(\ref{solution}). 

\setcounter{equation}{0}
\renewcommand{\theequation}{B\arabic{equation}}

\section{Generators of $S_4$}
\label{generators}
The generators of a finite group are not unique. For our purpose it is
useful to consider the Klein four-group, which is an Abelian subgroup of
$S_4$, given by $k_1 = (12)(34)$, $k_2 = (14)(23)$, 
$k_3 = (13)(24)$ and the unit
element. We further need one three-cycle, say $s = (123)$, and one
transposition, say $t = (12)$. All elements of $S_4$ can be obtained as
products of these permutations---see for instance~\cite{GL-review}.
However, what we are really interested in is a faithful three-dimensional 
irreducible representation of $S_4$. Here we display the ${\bf 3}$ as derived
in~\cite{GL-review}:
\begin{eqnarray} 
k_1 & \mapsto & S_1 = \left( \begin{array}{rrr}
1 & 0 & 0 \\ 0 & -1 & 0 \\ 0 & 0 & -1 
\end{array} \right), 
\nonumber \\
k_2 & \mapsto & S_2 = \left( \begin{array}{rrr}
-1 & 0 & 0 \\ 0 & 1 & 0 \\ 0 & 0 & -1 
\end{array} \right), 
\label{Si} \\
k_3 & \mapsto & S_3 = \left( \begin{array}{rrr}
-1 & 0 & 0 \\ 0 & -1 & 0 \\ 0 & 0 & 1 
\end{array} \right)
\nonumber
\end{eqnarray}
and 
\begin{equation}\label{EB}
s \mapsto E = \left( \begin{array}{rrr}
0 & 1 & 0 \\ 0 & 0 & 1 \\ 1 & 0 & 0 
\end{array} \right), 
\quad
t \mapsto B = \left( \begin{array}{rrr}
-1 & 0 & 0 \\ 0 & 0 & -1 \\ 0 & -1 & 0 
\end{array} \right).
\end{equation}
In the spirit of section~\ref{residual} we call this basis of the
${\bf 3}$ of $S_4$ basis~1. The second three-dimensional irreducible
representation, ${\bf 3'}$, 
is obtained is obtained from ${\bf 3}$ by a sign change in equation~(\ref{EB}),
namely $t \mapsto -B$. 

It is also useful to have the above generators in the basis where $E$ is
diagonal. With the similarity transformation 
$U_\omega^\dagger A U_\omega = \tilde A$ where $U_\omega$ is given by
equation~(\ref{Uomega}) we obtain (see for instance~\cite{merlo,ahn})
\begin{equation}
\tilde S_1 = \frac{1}{3} \left( \begin{array}{rrr}
-1 & 2 & 2 \\ 2 & -1 & 2 \\ 2 & 2 & -1 
\end{array} \right), 
\;\;
\tilde S_2 = \frac{1}{3} \left( \begin{array}{ccc}
-1 & 2\omega & 2\omega^2 \\ 2\omega^2 & -1 & 2\omega \\ 
2\omega & 2\omega^2 & -1 
\end{array} \right), 
\;\;
\tilde S_3 = \frac{1}{3} \left( \begin{array}{ccc}
-1 & 2\omega^2 & 2\omega \\ 2\omega & -1 & 2\omega^2 \\ 
2\omega^2 & 2\omega & -1 
\end{array} \right)
\end{equation}
and 
\begin{equation}
\tilde E = \diag \left( 1, \omega, \omega^2 \right),
\quad
\tilde B = B.
\end{equation}
In the spirit of section~\ref{residual} this is the ${\bf 3}$ given in
basis~2.

\end{document}